
\input harvmac.tex
\overfullrule=0pt

\def\simge{\mathrel{%
   \rlap{\raise 0.511ex \hbox{$>$}}{\lower 0.511ex \hbox{$\sim$}}}}
\def\simle{\mathrel{
   \rlap{\raise 0.511ex \hbox{$<$}}{\lower 0.511ex \hbox{$\sim$}}}}

\def\slashchar#1{\setbox0=\hbox{$#1$}           
   \dimen0=\wd0                                 
   \setbox1=\hbox{/} \dimen1=\wd1               
   \ifdim\dimen0>\dimen1                        
      \rlap{\hbox to \dimen0{\hfil/\hfil}}      
      #1                                        
   \else                                        
      \rlap{\hbox to \dimen1{\hfil$#1$\hfil}}   
      /                                         
   \fi}                                         %
\def\CH{{\cal H}}

\def\ts{\thinspace}
\def\ra{\rightarrow}

\def\ra{\rightarrow}
\def\ol{\bar}

\def\gev{{\rm GeV}}
\def\tev{{\rm TeV}}

\def\half{\textstyle{ { 1\over { 2 } }}}
\def\third{\textstyle{ { 1\over { 3 } }}}
\def\twothirds{\textstyle{ { 2\over { 3 } }}}
\def\sixth{\textstyle{ { 1\over { 6 } }}}
\def\fivesix{\textstyle{ { 5\over { 6 } }}}
\def\nin{\noindent}
\def\Gew{SU(2)\otimes U(1)}
\def\Gtc{G_{TC}}
\def\Getc{G_{ETC}}
\def\su{SU(3)}
\def\suc{SU(3)_C}
\def\uone{U(1)_1}
\def\utwo{U(1)_2}

\def\condt{\langle \bar T T\rangle}

\def\condtij{\langle \bar T^i_L T^j_R\rangle}
\def\conduij{\langle \bar U^i_L U^j_R\rangle}
\def\conddij{\langle \bar D^i_L D^j_R\rangle}
\def\condtbt{\langle \bar t t\rangle}

\def\Dzero{{\rm D}\slashchar{\rm O}}

\def\myfoot#1#2{{\baselineskip=14.4pt plus 0.3pt\footnote{#1}{#2}}}

\Title{\vbox{\baselineskip12pt\hbox{BUHEP--95--11}
\hbox{FERMILAB--PUB--95/052--T}
\hbox{hep-ph/9503433}}}
{Natural Topcolor--Assisted Technicolor}

\smallskip
\centerline{Kenneth Lane\myfoot{$^{\dag }$}{lane@buphyc.bu.edu}}
\smallskip\centerline{Department of Physics, Boston University}
\centerline{590 Commonwealth Avenue, Boston, MA 02215}
\centerline{and}
\smallskip
\centerline{Estia Eichten\myfoot{$^{\ddag }$}{eichten@fnal.gov}}
\smallskip\centerline{Fermi National Accelerator Laboratory}
\centerline{P.O.~Box 500 Batavia, IL 60510}
\vskip .3in

\centerline{\bf Abstract}

We construct a prototype of topcolor--assisted technicolor in which,
although both top and bottom quarks acquire some mass from extended
technicolor, strong $U(1)$ couplings of technifermions are isospin
symmetric and all gauge anomalies vanish. There is a mechanism for mixing
between the light and heavy generations and there need be no very light
pseudo--Goldstone bosons.

\bigskip

\Date{3/95}

\vfil\eject

Technicolor was invented to provide a natural, dynamical explanation for
electroweak symmetry breaking
\ref\tcref{S.~Weinberg, Phys.~Rev.~{\bf D19}, 1277 (1979)\semi
L.~Susskind, Phys.~Rev.~{\bf D20}, 2619 (1979).}.
Here, $\Gew$ is broken down to $U(1)_{EM}$ by technifermion condensates
$\condt$ generated by strong technicolor (TC) interactions. These
interactions have the characteristic energy scale $\Lambda_{TC}
\simeq \Lambda_{EW} \simeq 1\,\tev$. To account for
explicit breaking of quark and lepton flavor symmetries
within the spirit of technicolor, new gauge interactions, encompassing
technicolor and known as extended technicolor (ETC), also had to be
invented
\ref\sdlsetc{S.~Dimopoulos and L.~Susskind, Nucl.~Phys.~{\bf B155}, 237
(1979).},
\ref\eekletc{E.~Eichten and K.~Lane, Phys.~Lett.~{\bf 90B}, 125
(1980).}.
Unfortunately, there seems to be no natural way to
account for the extremely large mass, $m_t \simeq 175\,\gev$, of the top
quark
\ref\toprefs{F.~Abe, et al., The CDF Collaboration, Phys.~Rev.~Lett.~{\bf
73}, 225 (1994); Phys.~Rev.~{\bf D50}, 2966 (1994);
FERMILAB-PUB-95/022-E\semi
S.~Abachi, et al., The $\Dzero$ Collaboration, FERMILAB-PUB-95/028-E.}
within the ETC framework
\ref\setc{T.~Appelquist, M.~B.~Einhorn, T.~Takeuchi and
L.~C.~R.~Wijewardhana, Phys.~Lett.~{\bf 220B}, 223 (1989);
T.~Takeuchi, Phys.~Rev.~{\bf D40}, 2697 (1989)\semi
V.A.~Miransky and K.~Yamawaki, Mod.~Phys.~Lett.~{\bf A4},129 (1989)\semi
K.~Matumoto, Prog.~Theor.~Phys.~{\bf 81}, 277 (1989) \semi
R.~S.~Chivukula, A.~G.~Cohen and K.~Lane, Nucl.~Phys.~{\bf B343}, 554
(1990).}.

Topcolor was invented as a minimal dynamical scheme to reproduce the
simplicity of the one--doublet Higgs model {\it and} explain a very
large top--quark mass
\ref\topcref{C.~T. Hill, Phys.~Lett.~{\bf B266}, 419 (1991) \semi
S.~P.~Martin, Phys.~Rev.~{\bf D45}, 4283 (1992);
{\it ibid}~{\bf D46}, 2197 (1992); Nucl.~Phys.~{\bf B398}, 359 (1993)\semi
M.~Lindner and D.~Ross, Nucl.~Phys.~{\bf  B370}, 30 (1992)\semi
R.~B\"{o}nisch, Phys.~Lett.{\bf B268}, 394 (1991)\semi
C.~T.~Hill, D.~Kennedy, T.~Onogi, H.~L.~Yu, Phys.~Rev.~{\bf D47}, 2940
(1993).}.
Here, a large top--quark condensate, $\condtbt$, is formed by strong
interactions at the energy scale, $\Lambda_t \gg m_t \sim \Lambda_{EW}$
\ref\topcondref{Y.~Nambu, in {\it New Theories in Physics}, Proceedings of
the XI International Symposium on Elementary Particle Physics, Kazimierz,
Poland, 1988, edited by Z.~Adjuk, S.~Pokorski and A.~Trautmann (World
Scientific, Singapore, 1989); Enrico Fermi Institute Report EFI~89-08
(unpublished)\semi
V.~A.~Miransky, M.~Tanabashi and K.~Yamawaki, Phys.~Lett.~{\bf
221B}, 177 (1989); Mod.~Phys.~Lett.~{\bf A4}, 1043 (1989)\semi
W.~A.~Bardeen, C.~T.~~Hill and M.~Lindner, Phys.~Rev.~{\bf D41},
(1990).}.
In order that the resulting low--energy theory simulate the standard
model, this scale must be very high---$\Lambda_t \sim 10^{15}\,\gev$.
Unfortunately, the topcolor scenario is unnatural, requiring a fine--tuning
of couplings of order one part in $\Lambda_t^2/m_t^2 \simeq 10^{25}$.

Recently, Hill has proposed joining technicolor and topcolor
\ref\hill{C.T.~Hill, hep-ph/9411426, FERMILAB-PUB-94/395-T.}.
His idea is that electroweak symmetry breaking is driven mainly by
technicolor interactions strong near $1\,\tev$ and that light quark and
lepton masses are generated by ETC. In addition, topcolor interactions with
a scale also near $1\,\tev$ generate $\condtbt$ and the very large
top--quark mass. This neatly removes the objections that topcolor is
unnatural and that technicolor cannot generate a large top mass. In this
scenario, topcolor is an ordinary asymptotically free gauge theory, but it
is still necessary that technicolor be a walking gauge theory
\ref\wtc{B.~Holdom, Phys.~Rev.~{\bf D24}, 1441 (1981);
Phys.~Lett.~{\bf 150B}, 301 (1985)\semi
T.~Appelquist, D.~Karabali and L.~C.~R. Wijewardhana,
Phys.~Rev.~Lett.~{\bf 57}, 957 (1986)\semi
T.~Appelquist and L.~C.~R.~Wijewardhana, Phys.~Rev.~{\bf D36}, 568
(1987)\semi
K.~Yamawaki, M.~Bando and K.~Matumoto, Phys.~Rev.~Lett.~{\bf 56}, 1335
(1986) \semi
T.~Akiba and T.~Yanagida, Phys.~Lett.~{\bf 169B}, 432 (1986).}
to escape large flavor--changing neutral currents~\eekletc.

In detail, Hill's scheme depends on separate color and weak hypercharge
interactions for the third and for the first two generations of quarks and
leptons. For example, the (electroweak eigenstate) third generation
$(t,b)_{L,R}$ may transform with the usual quantum numbers under the gauge
group $\su_1 \otimes \uone$ while $(u,d)$, $(c,s)$ transform under a
separate group $\su_2 \otimes \utwo$. Topcolor is $\su_1$. Leptons of the
third and the first two generations transform in the obvious way to cancel
gauge anomalies. At a scale of order $1\,\tev$ (which may or may not be the
same as the electroweak scale), $\su_1 \otimes \su_2 \otimes \uone \otimes
\utwo$ is dynamically broken to the diagonal subgroup of ordinary color and
weak hypercharge, $\suc \otimes U(1)$. At this energy scale, the $\su_1
\otimes \uone$ couplings are strong while the $\su_2 \otimes \utwo$
couplings are weak. Top, but not bottom, condensation is driven by the fact
that the $\su_1 \otimes \uone$ interactions are supercritical for top
quarks, but subcritical for bottom.\foot{A large bottom condensate is not
generated because the topcolor $SU(3)$ symmetry is broken and the
interaction does not grow stronger as one descends to lower energies.} This
difference is caused by the $\uone$ couplings of $t$ and $b$. If this
topcolor--assisted technicolor (TC2) scenario is to be natural, i.e., no
fine--tuning of the $\su_1$ is required, the $\uone$ couplings {\it cannot}
be weak.

Chivukula, Dobrescu and Terning (CDT) have argued that the TC2 proposal
cannot be both natural and consistent with experimental measurements of the
parameter $\rho = M_W^2/M_Z^2 \cos^2\theta_W$
\ref\cdt{R.~S.~Chivukula, B.~A.~Dobrescu and J.~Terning, hep-ph/9503203,
BUHEP-95-8.}.
Their strongest criticism is that even degenerate up and down
technifermions are likely to have custodial--isospin violating couplings to
the strong $\uone$ and that this leads to large contributions to $\rho$. To
prevent this, CDT showed that the $\uone$ coupling must be so small that
it is necessary to tune the $\su_1$ coupling to within~1\% of the critical
value for top condensation {\it and} to increase the topcolor boson mass
above $4.5\,\tev$.

The argument presented by CDT that the technifermions' $\uone$ couplings
violate custodial isospin proceeds as follows: The $(t,b)$ chiral
symmetries must be broken explicitly by ETC interactions to avoid unwanted
massless bosons. In particular, part of $m_t$ must arise from ETC (see
Ref.~\hill). Thus, technifermions must couple to $\uone$. If ETC commutes
with electroweak $SU(2)$ and if $m_b$ also arises in part from
ETC,\foot{This may not be necessary if $\su_1$ instanton effects can
produce all of $m_b$; see~\hill.} then the right--handed technifermions
$U_R$ and $D_R$ to which $t$ and $b$ couple must have different $\uone$
couplings.

CDT further state that custodially--invariant $\uone$ couplings to
technifermions may be difficult to arrange because of the need to cancel
all gauge anomalies. The difficulty here is that, in cancelling the
anomalies with extra fermions, one must not introduce extra unbroken chiral
symmetries~\eekletc. Finally, they
stress that there must be mixing between the third and first two
generations and this further constrains hypercharge assignments.

In this Letter, we construct a prototype of TC2 that can overcome these
difficulties. In particular, provided that technifermion condensates align
properly:

\itemitem{1.)} Both $t$ and $b$ get some mass from ETC interactions.

\itemitem{2.)} $\uone$ couplings of technifermions preserve custodial
$SU(2)$.

\itemitem{3.)} All gauge anomalies vanish.

\itemitem{4.)} There is mixing between the third and first two generations.

\itemitem{5.)} The only spontaneously broken technifermion chiral symmetries
that are not also explicitly broken by ETC are the electroweak $SU(2)\otimes
U(1)$.

\nin Thus, the $\uone$ interaction can be moderately strong and the TC2
interactions natural.

Our prototype is incomplete in several ways: First, we do not specify the
ETC gauge group, $\Getc$; the existence of the desired ETC four--fermion
interactions is assumed. They are invariant under the $\su$ and $U(1)$
groups. For simplicity, we assume that $\Getc$ commutes with electroweak
$SU(2)$. Second, we do not specify how $\su_1 \otimes \su_2 \otimes \uone
\otimes \utwo$ is broken. An example of this was given in Ref.~\hill. We
might have to modify our $\uone \otimes \utwo$ charge assignments and
anomaly cancellations to accommodate this symmetry breaking. In our model,
there is $\uone \otimes \utwo$ breaking due to technifermion condensation.
We do not know yet whether this is sufficient. Third, we do not discuss
leptons other than to assume that they are paired with quarks to cancel
gauge anomalies in the usual way. Their masses may arise from ETC
interactions similar to those in Eq.~(1) below.

Our model has three doublets of technifermions, all of which are assumed to
transform according to the same complex irreducible representation of the
technicolor gauge group, $\Gtc$. They are also assumed to be singlets under
$\su_1 \otimes \su_2$.\foot{Thus, ETC bosons connecting these fermions to
quarks must be triplets under the appropriate $\su$ group.} The
technifermion doublets are denoted by $T^l_{L,R} = (U^l, D^l)_{L,R}$,
coupling to the first two (``light'') generation quarks via ETC; $T^t_{L,R}
= (U^t, D^t)_{L,R}$, giving the top quark its ETC--mass; and $T^b_{L,R} =
(U^b, D^b)_{L,R}$ giving the bottom quark its ETC--mass. To simplify our
presentation, we assume for now that condensates are flavor--diagonal:
$\conduij = \conddij \propto \delta_{ij}$ in the correct
chiral--perturbative vacuum
\ref\vacalign{R.~Dashen, Phys.~Rev.~{\bf D3}, 1879 (1971);
S.Weinberg, Phys.~Rev.~{\bf D13}, 974 (1976)\semi
E.~Eichten, K.~Lane and J.~Preskill, Phys.~Rev.~Lett.~{\bf 45}, 225
(1980)\semi
M.~Peskin, Nucl~Phys.~{\bf B175}, 197 (1980);
J.~Preskill, Nucl.~Phys.~{\bf B177}, 21 (1981).}.
Below, when we discuss vacuum alignment, we shall see that
these condensates are matrices in flavor space. We shall find that,
generically, they still induce nonzero quark masses, but the
precise outcome depends on the details of ETC symmetry and its breaking.

To generate light and heavy quark masses, we assume that ETC interactions
produce couplings of the following form:
\eqn\qTTq{\eqalign{
&\CH_{\ol u_i u_j} = {g^2_{ETC} \over {M^2_{ETC}}} \ts \ol q^l_{iL}
\gamma^\mu T^l_L \ts \ol U^l_R \gamma_\mu u_{jR}  \ts + \ts\ts {\rm h.c.} \cr
&\CH_{\ol d_i d_j} = {g^2_{ETC} \over {M^2_{ETC}}} \ts \ol q^l_{iL}
\gamma^\mu T^l_L \ts \ol D^l_R \gamma_\mu d_{jR}  \ts + \ts\ts {\rm h.c.} \cr
&\CH_{\ol t t} = {g^2_{ETC} \over {M^2_{ETC}}} \ts \ol q^h_L \gamma^\mu T^t_L
\ts \ol U^t_R \gamma_\mu t_R   \ts + \ts\ts {\rm h.c.} \cr
&\CH_{\ol b b} = {g^2_{ETC} \over {M^2_{ETC}}} \ts \ol q^h_L \gamma^\mu T^b_L
\ts \ol D^b_R \gamma_\mu b_R   \ts + \ts\ts {\rm h.c.} \cr}}
Here, $g_{ETC}$ and $M_{ETC}$ stand for generic ETC couplings and gauge
boson mass matrices; also, $q^l_{iL} = (u_i, d_i)_L$ for $i=1,2$ and $q^h_L
= (t,b)_L$. The assignments of the $\uone$ and $\utwo$ hypercharges, $Y_1$
and $Y_2$, for the quarks and those for technifermions consistent with
these interactions and with electric charge, $Q = T_3 + Y_1 +Y_2$, are listed
in Table~1 in terms of six parameters ($x_{1,2}$, $y_{1,2}$, $z_{1,2}$) to
be determined. The strong $\uone$ couplings of the right and left--handed
technifermions are isospin symmetric. This is possible because {\it
different} technifermions give mass to $t$ and $b$.

The conditions that gauge anomalies vanish are:
\eqn\anomaly{\eqalign{
 \uone [SU(2)_{EW}]^2: &\qquad x_1 + y_1 + z_1 = 0 \cr
 \utwo [SU(2)_{EW}]^2: &\qquad x_2 + y_2 + z_2 = 0 \cr
 [\uone]^3: &\qquad x_1(y_1 - z_1 + \half) = 0 \cr
 [\utwo]^3: &\qquad x_2(y_2 - z_2 - \half) = 0 \cr
 [\uone]^2\utwo: &\qquad x_1 - x_2 + 4(y_1y_2 - z_1 z_2) = 0 \cr}}
The $U(1)_i[\Gtc]^2$ anomaly conditions are automatically satisfied by the
hypercharge assignments in Table~1. Earlier anomaly conditions in
Eqs.~\anomaly\ were imposed on the later ones. Thus, the $[\utwo]^2\uone$
condition is the same as for $[\uone]^2\utwo$.

To choose among the solutions to Eqs.~\anomaly, we insist that there
mixing between the third and first two generations.
Specifically, we require that there exist ETC--generated
four--technifermion (4T) interactions which connect $T^l$ to $T^t$ or to
$T^b$ and which are consistent with $SU(2)\otimes \uone\otimes\utwo$. Then,
once color and hypercharge symmetries break to $\suc \otimes U(1)$, these
operators can mix the heavy and light generations. Still assuming $\condtij
\propto \delta_{ij}$, there are four possible 4T operators: $\ol
T^l_L\gamma^\mu T^t_L \ts \ol D^t_R \gamma_\mu D^b_R,\ts$ $\ol
T^b_L\gamma^\mu T^t_L \ts \ol D^t_R \gamma_\mu D^l_R,\ts$, $\ol
T^l_L\gamma^\mu T^b_L \ts \ol U^b_R \gamma_\mu U^t_R,\ts$
and $\ol
T^t_L\gamma^\mu T^b_L \ts \ol U^b_R \gamma_\mu U^l_R$. These have the
potential to induce the mixings $b_R$--$s_L, d_L$; $\ts b_L$--$s_R, d_R$;
$\ts t_R$--$c_L, u_L$; and $t_L$--$u_R, c_R$, respectively.

The known mixing between the third and the first two generations is in the
Kobayashi--Maskawa matrix for left--handed quarks. It is $|V_{cb}| \simeq
|V_{ts}| \simeq$~0.03--0.05 $\sim m_s/m_b$ and $|V_{ub}| \simeq |V_{td}|
\simeq $~0.002--0.015 $\sim \sin\theta_C \ts m_s/m_b$
\ref\pdg{Particle Data Group, Phys.~Rev.~{\bf D50}, 1174 (1994).}.
A nonzero term~$\delta m \sim m_s$ in the $\ol s_L b_R$ element of the
quark mass matrix is needed to produce mixing of this magnitude. Thus, only
the first of the 4T operators above has the correct flavor and chiral
structure. Requiring this operator leads to two solutions to the anomaly
conditions, which we call cases~A and B. The case~A solution is:
\eqn\solA{
{\rm Case A:}\qquad
x_1 = -\half , \ts\ts y_1 = 0, \ts\ts z_1 = \half; \quad
x_2 = \half , \ts\ts y_2 = 0, \ts\ts z_2 = -\half \ts.}
The 4T operators allowed by these hypercharges (and that influence the
vacuum's alignment) are $\CH_{\ol l t \ol t b}$, $\CH_{\ol t b \ol b l}$,
$\CH_{\ol b l \ol l t}$,  $\CH_{\ol t l \ol l b}$, $\CH_{\ol t b \ol t b}$
and $\CH_{\rm diag}$, where, for example,
\eqn\opA{\eqalign{
&\CH_{\ol l t \ol t b}= {g^2_{ETC} \over {M^2_{ETC}}} \ts \ol T^l_L
 \gamma^\mu T^t_L \ts (a_{_U} \ol U^t_R \gamma_\mu U^b_R  + a_{_D} \ol
 D^t_R \gamma_\mu D^b_R)  \ts + \ts {\rm h.c.} \cr
&\CH_{{\rm diag}}= {g^2_{ETC} \over {M^2_{ETC}}} \ts \ol T^i_L
 \gamma^\mu T^i_L \ts (b_{_U} \ol U^j_R \gamma_\mu U^j_R  + b_{_D} \ol
 D^j_R \gamma_\mu D^j_R)  \ts . \cr }}
The constants $a_{_{U,D}}$,... stand for unknown ETC--model--dependent
factors and, in the diagonal interaction, $i,j = l,t,b$.
\foot{The need for custodial isospin violation in these operators was
discussed in Ref.~\eekletc. Since the operators generate technifermion
``hard'' masses of at most a few GeV, they are not expected to contribute
excessively to $\rho-1$; see Ref~\ref\taref{T.~Appelquist, et al.,
Phys.~Rev.~Lett.~{\bf 53}, 1523 (1984); Phys.~Rev.~{\bf D31}, 1676 (1985).}.}
These flavor--diagonal interactions may arise, for example, from broken
$U(1)$ subgroups of $\Getc$. The case~B solution to the anomaly conditions
are
\eqn\solB{
{\rm Case B:}\qquad
x_1 = 0 , \ts\ts y_1 = -1, \ts\ts z_1 = 1; \quad
x_2 = 0 , \ts\ts y_2 = 1, \ts\ts z_2 = -1 \ts.}
The allowed operators are $\CH_{\ol l t \ol t b}$, $\CH_{\ol l b \ol b t}$
and $\CH_{\rm diag}$.

We obtain a constraint on $b$--$s$ mixing as follows: As noted above, the
transition $b_R \ra D^b_R \ra D^l_L \ra s_L$ requires both the interaction
$\CH_{\ol l t \ol t b}$ {\it and} breaking of the separate color and
hypercharge groups to $\suc\otimes U(1)$. Since the technifermions in our
prototype are $\su_1 \otimes \su_2$ singlets, there must be operator
effecting this breaking in the ETC boson mass matrix. This operator
transforms as $(\ol {\bf 3}, {\bf 3}; \fivesix, -\fivesix)$ in cases~A,B.
Let us denote the corresponding mass--mixing term by $\delta M^2_{ETC}$.
Also, denote by $M_s$ and $M_b$ the ETC boson masses that generate $m_s$
and the ETC contribution to the $b$--quark mass, $m_b^{ETC}$. We expect
$M_s \simge 100\,\tev$ and $M_s/M_b \sim m_b^{ETC}/m_s$ in a walking
technicolor model. Then, we estimate that
\eqn\mratio{
{\delta m \over {m_b^{ETC}}} \sim {\delta M^2_{ETC} \over {M^2_s}} \ts\ts.}
If topcolor instanton and ETC contributions to $m_b$ add, then $\delta m
/m_b \simle  \delta M^2_{ETC}/M^2_s$. The quark masses in Eq.~\mratio\ are
renormalized at the ETC scale $M_b$, which is above the scale at which
$\su_1 \otimes \su_2 \otimes \uone \otimes \utwo$ is broken. {\it If} the
the effect of renormalization down to about $1\,\tev$ of the operators
involved in this ratio is small, then $|V_{cb}| \simeq |V_{ts}| \simle
\delta M^2_{ETC}/ M^2_s$. This requires $\delta M_{ETC} \sim 10\,\tev$, too
large to be compatible with a topcolor breaking scale near $1\,\tev$.
Obviously, the issue of renormalization of the mixing parameters down to
the QCD scale must be addressed in a more complete model. Obtaining mixing
of the right magnitude will be a challenge.

Turn now to the question of vacuum alignment. If broken ETC and $U(1)$
interactions of technifermions may be treated as a perturbation, the flavor
symmetry group of the technicolor sector is $G_\chi = SU(6)_L \otimes
SU(6)_R$. When TC interactions become strong, $G_\chi$ breaks spontaneously
to an $SU(6)$ subgroup. For simplicity, we have assumed that this is the
diagonal subgroup, $S_\chi = SU(6)_V$. This pattern of symmetry breaking
assumes that the ground state minimizing the expectation value of the
chiral symmetry breaking interactions is characterized by the
flavor--diagonal condensates\foot{On dimensional grounds, we expect
$\Delta_T \simeq 4 \pi F_T^3$
\ref\dimnal{A.~Manohar and H.~Georgi, Nucl.~Phys.~{\bf B234}, 189
(1984)\semi H.~Georgi and L.~Randall, Nucl.~Phys.~{\bf B276}, 241
(1986).},
where $F_T \simeq 246\,\gev/\sqrt{3}$, consistent with the $SU(6)\otimes
SU(6)$ chiral symmetry of technifermions.}
\eqn\tconds{\conduij = \conddij = -\half  \Delta_T \ts \delta_{ij}\ts,
\quad (i,j = l,t,b) \ts.}

Whether this or another pattern is preferred depends on the relative
strengths and signs of the explicit $G_\chi$--breaking interactions in
Eqs.~\qTTq\ and the 4T operators and strong $\uone$ interactions in cases~A
or~B. For example, the $\uone$ interactions of case~A prefer the
condensates to align as $\langle \ol T_L^l T_R^l \rangle = \langle \ol
T_L^t T_R^b \rangle = \langle \ol T_L^b T_R^t \rangle = -\half \Delta_T$,
while those in case~B prefer the diagonal alignment in Eq.~\tconds. If
the former alignment occurred, it would not be possible to generate proper
ETC masses for the $t$ and $b$ quarks. In a TC theory whose coupling
evolves very little below $M_{ETC}$, the three types of interaction make
contributions to the vacuum energy which are nominally of order $m_t^{ETC}
\condtbt$, $g^2_{ETC} \Delta_T^2/\Lambda_T^2$ (where $\Lambda_T \simeq {\rm
few} \times F_T$), and $g^2_{\uone} \ts F_T^4$. The $\uone$ coupling is
strong, but so is $g^2_{ETC}$ in a walking gauge theory
\ref\multi{K.~Lane and E.~Eichten, Phys.~Lett.~{\bf 222B} (1989)~274 \semi
K.~Lane and M~V.~Ramana, Phys.~Rev.~{\bf D44} (1991)~2678.}.
Therefore, it seems likely that the ETC interactions will be the decisive
ones. In any case, it is easy to see that the ETC and $\uone$ interactions
in either case explicitly violate all spontaneously broken
chiral symmetries except for the electroweak ones. Thus, there are no light
Goldstone bosons left over.

A full discussion of vacuum alignment is not possible in this Letter. We
need to construct definite ETC models and determine the allowed chiral
symmetry breaking interactions and their strengths before we can state what
vacuum alignment patterns occur and whether they produce the desired quark
and lepton masses. In lieu of that, we briefly summarize the results of a
study for our case~B choice of hypercharges.

For simplicity, consider only the allowed ETC interactions. If only the
interaction $\CH_{\ol l t \ol t b}$ exists and has the sign given indicated
in Eq.~\opA, the condensates align as $\langle \ol T_L^t T_R^t \rangle =
\langle \ol T_L^l T_R^b \rangle = \langle \ol T_L^b T_R^l \rangle = -\half
\Delta_T$. If only $\CH_{\ol l b \ol b t}$ exists, interchange $t$ and~$b$.
If both interactions $\CH_{\ol l t \ol t b}$  and $\CH_{\ol l b \ol b t}$
have positive signs, the one with the dominant coefficient determines the
condensation pattern. If they have the same strength, either pattern
minimizes the vacuum energy, so that the vacuum is doubly degenerate. If
the diagonal ETC interactions $\CH_{\ol t t \ol t t}$ and $\CH_{\ol b b \ol
b b}$ also appear with comparable strength (or if the $\uone$ interactions
are as strong), the condensates in the correct vacuum form fully mixed
matrices in techniflavor space. It is clear that there is a broad range of
possibilities for vacuum alignment and that phenomenologically interesting
patterns can arise quite naturally.

Much work remains to construct a satisfactory model of topcolor--assisted
technicolor. One important issue is topcolor breaking. It is easy to break
topcolor using spectator fermions that introduce no gauge anomalies nor
unwanted Goldstone bosons. However, we believe it is preferable to
incorporate the breaking of topcolor with that of electroweak symmetry. An
even more ambitious program is to construct an ETC model, based on an
assumed pattern of symmetry breaking of some $\Getc$, and to complete the
vacuum alignment analysis. We are hopeful that progress can be made on
these issues.

\bigskip

We thank Sekhar Chivukula, Bogdan Dobrescu, Howard Georgi and Chris Hill
for discussions and comments. KL's research is supported in part by the
Department of Energy under Grant~No.~DE--FG02--91ER40676. EE's research is
supported by the Fermi National Accelerator Laboratory, which is operated
by Universities Research Association, Inc., under
Contract~No.~DE--AC02--76CHO3000.

\listrefs

\centerline{\vbox{\offinterlineskip
\hrule\hrule
\halign{&\vrule#&
  \strut\quad#\hfil\quad\cr
height4pt&\omit&&\omit&&\omit&&\omit&\cr\cr
&\hfill Particle \hfill&&\hfill$Y_1$ \hfill&&\hfill
 $Y_2$\hfill&&\hfill$Q = T_3 + Y_1 + Y_2$ \hfill &\cr\cr
height4pt&\omit&&\omit&&\omit&&\omit&\cr\cr
\noalign{\hrule\hrule}
height4pt&\omit&&\omit&&\omit&&\omit&\cr\cr
&$q_L^l$&&\hfill$0$\hfill&&\hfill$\sixth$\hfill
&&\hfill$\twothirds$, $-\third$\hfill&\cr\cr
\noalign{\hrule}
height4pt&\omit&&\omit&&\omit&&\omit&\cr\cr
&$c_R$, $u_R$&&\hfill$0$\hfill&&\hfill$\twothirds$\hfill
&&\hfill$\twothirds$\hfill&\cr\cr
\noalign{\hrule}
height4pt&\omit&&\omit&&\omit&&\omit&\cr\cr
&$d_R$, $s_R$&&\hfill$0$\hfill&&\hfill$-\third$\hfill
&&\hfill$-\third$\hfill&\cr\cr
\noalign{\hrule}
height4pt&\omit&&\omit&&\omit&&\omit&\cr\cr
&$q_L^h$&&\hfill$\sixth$\hfill&&\hfill$0$\hfill
&&\hfill$\twothirds$, $-\third$\hfill&\cr\cr
\noalign{\hrule}
height4pt&\omit&&\omit&&\omit&&\omit&\cr\cr
&$t_R$&&\hfill$\twothirds$\hfill&&\hfill$0$\hfill
&&\hfill$\twothirds$\hfill&\cr\cr
\noalign{\hrule}
height4pt&\omit&&\omit&&\omit&&\omit&\cr\cr
&$b_R$&&\hfill$-\third$\hfill&&\hfill$0$\hfill
&&\hfill$-\third$\hfill&\cr\cr
\noalign{\hrule}
height4pt&\omit&&\omit&&\omit&&\omit&\cr\cr
&$T_L^l$&&\hfill$x_1$\hfill&&\hfill$x_2$\hfill
&&\hfill$\pm \half + x_1 + x_2$\hfill&\cr\cr
\noalign{\hrule}
height4pt&\omit&&\omit&&\omit&&\omit&\cr\cr
&$U_R^l$&&\hfill$x_1$\hfill&&\hfill$x_2 + \half$\hfill
&&\hfill$\half + x_1 + x_2$\hfill&\cr\cr
\noalign{\hrule}
height4pt&\omit&&\omit&&\omit&&\omit&\cr\cr
&$D_R^l$&&\hfill$x_1$\hfill&&\hfill$x_2 - \half$\hfill
&&\hfill$-\half + x_1 + x_2$\hfill&\cr\cr
\noalign{\hrule}
height4pt&\omit&&\omit&&\omit&&\omit&\cr\cr
&$T_L^t$&&\hfill$y_1$\hfill&&\hfill$y_2$\hfill
&&\hfill$\pm \half + y_1 + y_2$\hfill&\cr\cr
\noalign{\hrule}
height4pt&\omit&&\omit&&\omit&&\omit&\cr\cr
&$U_R^t$&&\hfill$y_1 + \half$\hfill&&\hfill$y_2$\hfill
&&\hfill$ \half + y_1 + y_2$\hfill&\cr\cr
\noalign{\hrule}
height4pt&\omit&&\omit&&\omit&&\omit&\cr\cr
&$D_R^t$&&\hfill$y_1 + \half$\hfill&&\hfill$y_2 - 1$\hfill
&&\hfill$-\half + y_1 + y_2$\hfill&\cr\cr
\noalign{\hrule}
height4pt&\omit&&\omit&&\omit&&\omit&\cr\cr
&$T_L^b$&&\hfill$z_1$\hfill&&\hfill$z_2$\hfill
&&\hfill$\pm \half + z_1 + z_2$\hfill&\cr\cr
\noalign{\hrule}
height4pt&\omit&&\omit&&\omit&&\omit&\cr\cr
&$U_R^b$&&\hfill$z_1 - \half$\hfill&&\hfill$z_2 + 1$\hfill
&&\hfill$ \half + z_1 + z_2$\hfill&\cr\cr
\noalign{\hrule}
height4pt&\omit&&\omit&&\omit&&\omit&\cr\cr
&$D_R^b$&&\hfill$z_1 - \half$\hfill&&\hfill$z_2$\hfill
&&\hfill$-\half + z_1 + z_2$\hfill&\cr\cr
height4pt&\omit&&\omit&&\omit&&\omit&\cr\cr}
\hrule\hrule}}
\bigskip\bigskip
\noindent{TABLE 1: Quark and technifermion hypercharges and electric
charges. The parameters $x_i, y_i, z_i$ are determined in the text.}

\vfil\eject

\bye